\documentclass{article}
\usepackage{ijcai24}

\usepackage{times}
\usepackage{soul}
\usepackage{url}
\usepackage[hidelinks]{hyperref}
\usepackage[utf8]{inputenc}
\usepackage[small]{caption}
\usepackage{graphicx}
\usepackage{amsmath}
\usepackage{amsthm}
\usepackage{booktabs}
\usepackage{algorithm}
\usepackage{algorithmic}
\usepackage[switch]{lineno}
\usepackage{tabularx} 

\title{Interaction as Explanation: A User Interaction-based Method for Explaining Image Classification Models}

\author{
    Hyeonggeun Yun
    \affiliations
    Companoid Labs
    \emails
    hg.russ.yun@companoid.io
}

\begin{document}

\maketitle

\begin{abstract}
    In computer vision, explainable AI (xAI) methods seek to mitigate the 'black-box' problem by making the decision-making process of deep learning models more interpretable and transparent. Traditional xAI methods concentrate on visualizing input features that influence model predictions, providing insights primarily suited for experts. In this work, we present an interaction-based xAI method that enhances user comprehension of image classification models through their interaction. Thus, we developed a web-based prototype allowing users to modify images via painting and erasing, thereby observing changes in classification results. Our approach enables users to discern critical features influencing the model's decision-making process, aligning their mental models with the model's logic. Experiments conducted with five images demonstrate the potential of the method to reveal feature importance through user interaction. Our work contributes a novel perspective to xAI by centering on end-user engagement and understanding, paving the way for more intuitive and accessible explainability in AI systems.
\end{abstract}

\section{Introduction}
\label{sec:intro}


Significant advancements have been achieved across multiple disciplines based on deep neural network models. Notably, models have evolved substantially in computer vision to address several tasks, such as image classification, object recognition, and semantic segmentation. From convolutional neural network (CNN) to Transformer-based models, the computer vision models have demonstrated high performance on various datasets and tasks \cite{arkin2021survey,khan2022transformers,elngar2021image}. The practical applications of these models are extensive and diverse, including autonomous driving, optical character recognition, medical diagnostic imaging, and commercial product recommendation.

Despite the considerable advancements of deep neural networks, their deployment in the real world is fraught with challenges. Among these, the 'black-box' problem is a particularly prominent challenge \cite{yun2023driving,von2021transparency}. This issue refers to the opacity and interpretability of deep neural network-based models, wherein the algorithms and operational principles lack transparency, making it difficult for users to comprehend the model's decision-making process. The black-box problem can engender unpredictability in the outcomes generated by the models, thereby potentially affecting users' trust \cite{jiang2023situation}. Therefore, it is imperative to devise solutions addressing the black-box problem to apply models across diverse sectors successfully.

Thus, prior research has proposed explainable AI (xAI) methods to mitigate the black-box problem within computer vision. The representative method is a feature-based explanation method, highlighting the input features that affect the models' output and decision \cite{dwivedi2023explainable,liao2021human,kim2018interpretability,yeh2018representer,ribeiro2016should,selvaraju2017grad,mothilal2020explaining}. For instance, the Local Interpretable Model-Agnostic Explanations (LIME) technique identifies areas within input images that influence the outputs of image classification models \cite{ribeiro2016should}. Similarly, the gradient-weighted Class Activation Mapping (Grad-CAM) technique produces saliency maps to visualize important features based on gradients from the last layer of CNN models \cite{selvaraju2017grad}. These techniques provide intuitive visual explanations for better model understanding. Other feature-based methods include the Testing with Concept Activation Vectors (TCAV) technique, partial dependence plots (PDP), and individual conditional expectation (ICE) methods. TCAV technique measures the impact of concepts from input images on the model's output \cite{kim2018interpretability}. The partial dependence plots (PDP) and individual conditional expectation (ICE) methods illustrate how changes in inputs influence model outputs using a plot \cite{dwivedi2023explainable,liao2021human,khater2023skin,goldstein2015peeking}. These techniques are instrumental in highlighting the most significant features of the inputs. These xAI methods could be divided based on several criteria \cite{liao2021human,dwivedi2023explainable,kim2023help}, such as the representation of explanations (e.g., heatmap, concepts, plots), model dependency (model-specific vs. model-agnostic), and the scope of explanation (global vs. local). 


\begin{figure*}
  \centering
   \includegraphics[width=0.96\linewidth]{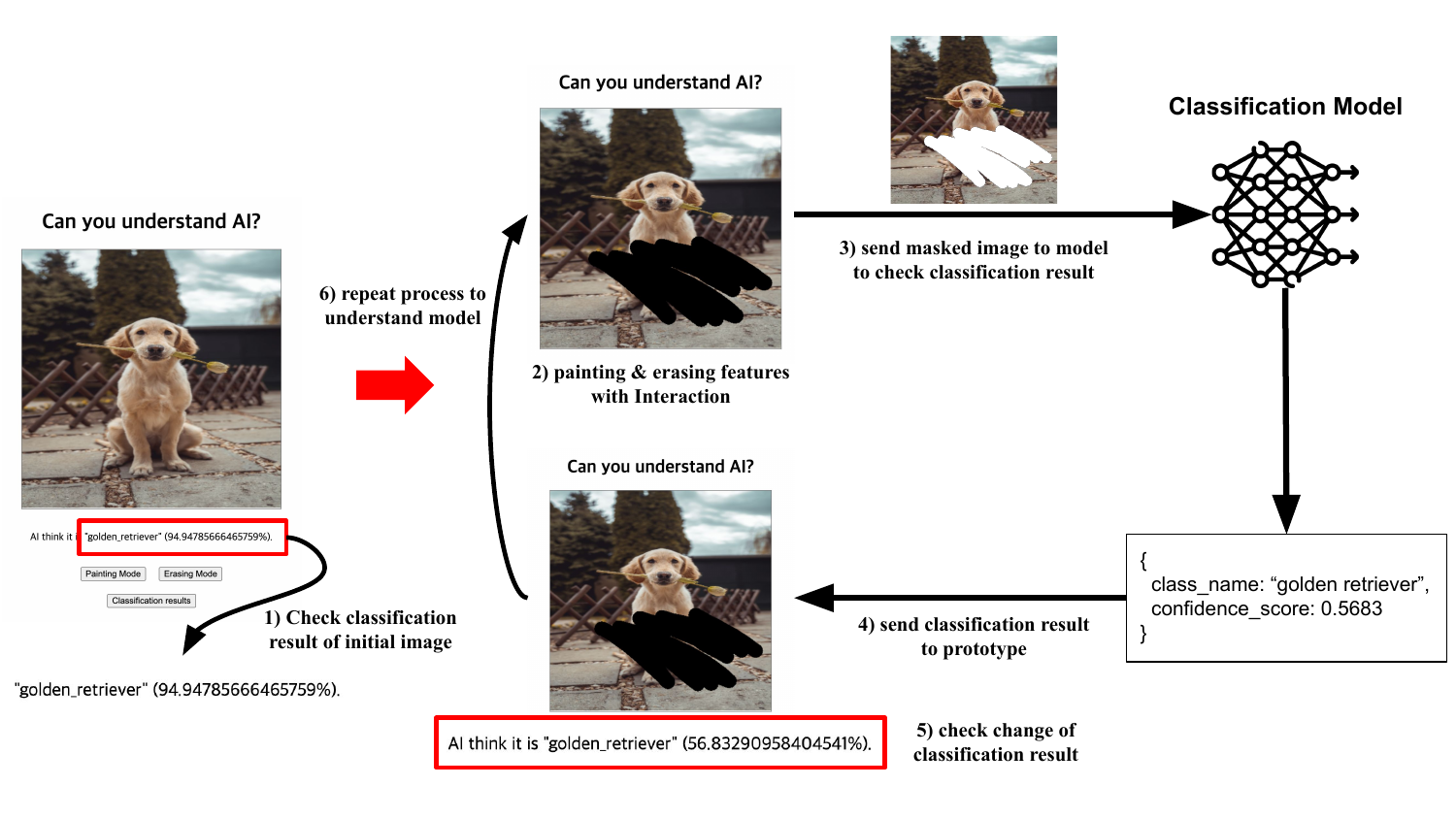}

   \caption{An example scenario of the interaction-based xAI method. 1) Our prototype displays the classification result of the initial image. 2) Users generate a masked image through the interaction. 3) The masked image is sent to the classification model to check the classification results. 4) The model sends the changed classification result back to the prototype. 5) The users could review the changed classification results. 6) The users could repeat the interaction process to deepen their understanding of the model through these interactions.}
   \label{fig:scenario}
\end{figure*}

The xAI methods play a crucial role by enabling users to comprehend how models process inputs \cite{patricio2023explainable,dwivedi2023explainable,liao2021human,recio2021case}. Furthermore, depending on the nature of the model and the specifics of the explanation needed, model designers could implement the appropriate xAI technique. However, previous investigations into xAI methods have primarily focused on explanations for AI experts \cite{kim2023help}. While end-users also require explainability and interpretability on models, current xAI methods may not fully address their requirements \cite{narwaria2022does}. The end-users want to participate in the design of xAI methods \cite{kim2023help}; however, these methods are currently developed without incorporating end-user feedback. Users frequently have xAI-related questions, such as "Why did the system produce this result?", "Which features are significant?", and "Under what circumstances do the results change?" \cite{liao2020questioning}.  However, existing methods often struggle to provide intuitive answers to these questions. To address this challenge, previous research has proposed xAI interfaces that facilitate iterative user interaction \cite{shneiderman2020bridging,weld2019challenge}. These interfaces have been shown to enhance user satisfaction and comprehension of the system, while also facilitating active exploration \cite{shneiderman2020bridging,du2019eventaction}. 

We propose an interaction-based xAI method that enables users to understand models by observing changes in image classification results through painting-based interactions. This approach allows users to easily express their xAI questions by directly masking areas of interest. Furthermore, our approach enables users to immediately verify differences in classification results, thereby intuitively inferring answers to their questions. For instance, in the context of medical image-based diagnostic classification tasks, traditional methods typically represent the importance of individual image pixels through visualizations such as heat maps. However, these representations may not be readily comprehensible to patients or medical experts. In contrast, our interaction-based method allows experts or patients to mask suspected tumor regions and directly observe the resultant changes in diagnostic classification.



To facilitate the interaction-based xAI method, we have developed a web-based prototype. This prototype features a canvas that displays an image of a specific object, allowing users to paint over the image directly on the canvas. Initially, users examine the classification results of the image—including the class and confidence score— before it has been painted. Subsequently, they paint over areas that they believe do not influence the classification results. Then, the painted parts of the image are masked and removed from the image classification. After painting, they check how the classification results of the masked image differ from the initial classification results. By repeating this process, users can identify critical image features that influence the model's input-to-output process and verify if their mental model is consistent with the model's classification criteria. This approach can be universally applied to any classification model via our prototype.

\section{Interaction-based xAI}
\label{sec:method}

The interaction-based xAI method enables users to understand image classification models with their painting interaction in a web-based prototype. Thus, we first implemented a web-based prototype that allows users to paint an image and provides the model's image classification results. Then, we described the interaction scenario of users to understand the models through the prototype and painting interaction.

\subsection{Prototype for Interaction-based xAI Method}

Our prototype was developed as a web-based system. We utilized the React library for frontend development and the FastAPI framework for backend development. The Pytorch was used for the deep learning framework \cite{steiner2019pytorch}. Our prototype consists of a canvas with an image of an object, a section showing the classification results along with the confidence score, buttons for painting mode and erasing mode, and a button to check the classification results.

The canvas of the prototype initially loads an image with a specific object from an image API. Users can paint unimportant features or erase painted areas on the canvas via swipe or touch interaction. Then, the painted area is masked, and the image is transmitted to the backend server for image classification. 

The section displaying the classification results and confidence score presents the class name along with the confidence score from the backend server. This enables users to review the classification results of the current masked image and how much confidence the model has in these results. The confidence score is expressed as a percentage.  

The buttons provide functions for the prototype. Buttons for painting and erasing modes allow users to paint and erase the image through interactions. The button for classification results masks the image's painted area and requests the masked image's classification results.

The backend server runs the model's inference on the image received from the frontend and responds to the request of the frontend. Thus, we first define an image classification model. The ResNet-50 model \cite{he2016deep}, pre-trained on the ImageNet dataset \cite{deng2009imagenet}, is employed as the model. Then, the server preprocesses the image; it is resized and normalized according to the ImageNet dataset standards. After that, the model performs the inference on the image. Finally, the server applies the softmax function on the model's output and sends the classification result --corresponding to the image class with the highest confidence-- back to the frontend.

\subsection{Scenario of Interaction-based xAI Method} \label{sec:scenario}

Through our prototype, we anticipate that users can understand the input features that affect the model's output. Furthermore, we expect them to be able to compare their initial understanding of the model to their understanding after using the prototype. This requires iterative masked image modification based on classification results. An illustrative example of the interaction-based xAI method is depicted in Figure \ref{fig:scenario}. The details are outlined below.

\begin{enumerate}
  \item Users first examine the image and the classification results (class name and confidence score) of the image.
  \item The users click a painting mode button and create a masked image by painting the features that they think do not affect the classification results through touch or swipe interaction.
  \item After painting, the users check the classification results of the masked image. 
  \item Based on the classification results, the users modify the masked image using painting or erasing modes as needed.
  \item The users can repeat the above process to understand which input features affect the model's output. 
\end{enumerate}

\section{Experiments}

\begin{table*}
  \centering
  \begin{tabularx}{\textwidth}{@{}lXX@{}}
    \toprule
    Image & First Interaction & Second Interaction \\
    \midrule
    golden retriever & masking background except dog & masking all except face \\ 
    soccer ball & masking background except ball & masking all except ball and leg\\
    coffee mug & masking background except mug & masking background and handle of mug\\
    bakery & masking others except for shelves & masking others except for shelves, a person, and some interior\\
    cinema & masking others except for screen and stage & masking others except for screen, stage, and some seats\\
    \bottomrule
  \end{tabularx}
  \caption{Details of interactions for the experiments. The table describes the parts masked in each interaction.}
  \label{tab:exp_design}
\end{table*}

We conducted experiments on five images to validate how the classification results and confidence scores are changed when our interaction-based xAI method is used to generate masked images. The five images were selected to include animals, objects, and places: a golden retriever, a soccer ball, a coffee mug, a bakery, and a cinema.

In the experiments, we painted the image by interaction with the scenario based on Section \ref{sec:scenario}. During the experiments, we generated two masked images with two separate interactions. In the first interaction, we painted all but the most representative part of the object. For animals and objects, we painted the parts that excluded the main subjects. For places, we painted everything except the elements most representative of each place. Subsequently, based on the results of the first interaction, we modified some parts of the image during the second interaction, either by painting or erasing. More details are provided in Table \ref{tab:exp_design}. 



\subsection{Experiments Results}

\begin{figure}
  \centering
   \includegraphics[width=0.98\linewidth]{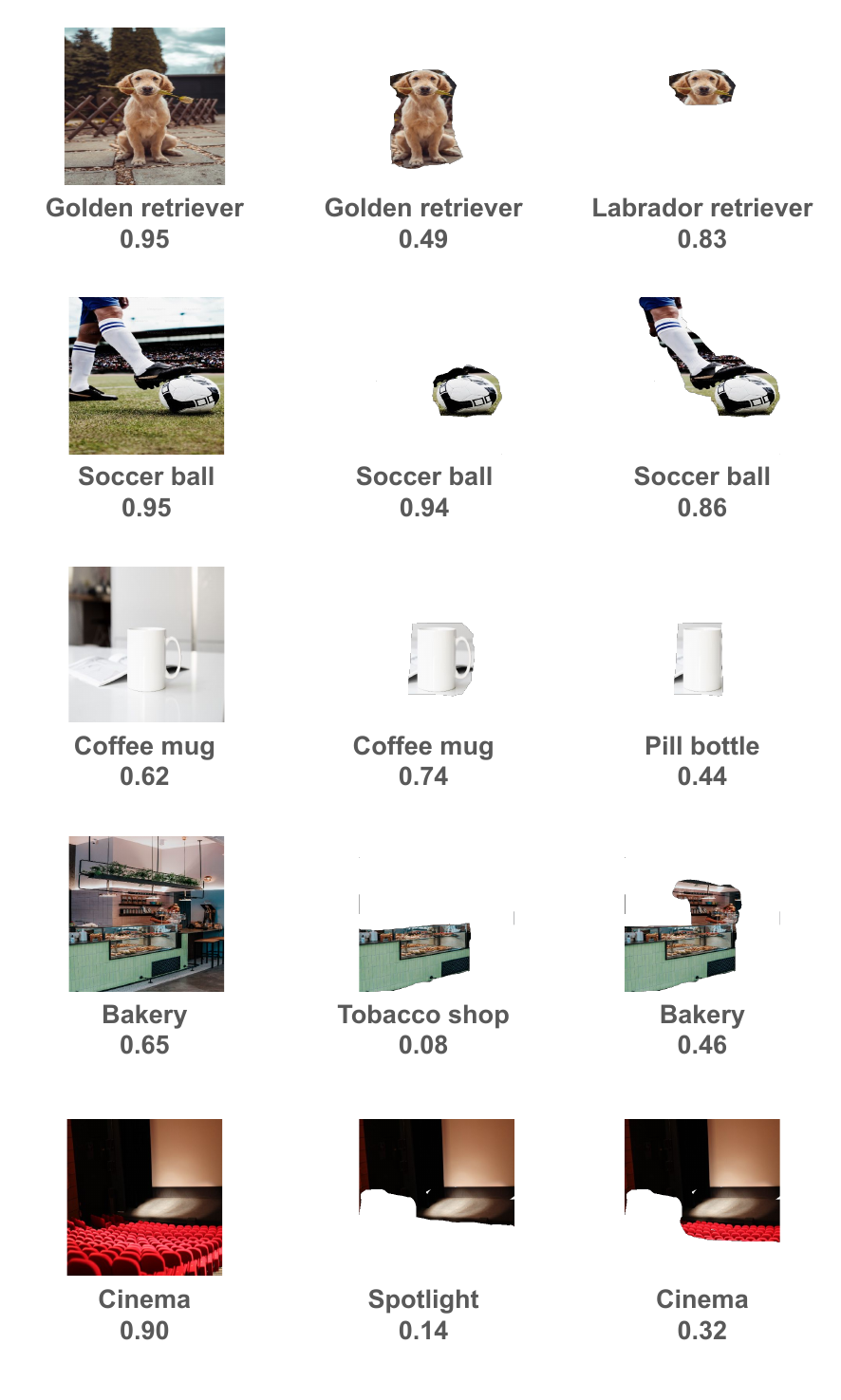}

   \caption{Experiment results of five images. The first column shows initial images. The second column shows images after the first interaction. The third column shows images after the second interaction. The classification results are shown below each image.}
   \label{fig:exp_results}
\end{figure}

The masked images after the interactions and classification results (class names and confidence scores) are shown in Figure \ref{fig:exp_results}.

For the golden retriever image, the background was masked during the first interaction. The classification result remained as a golden retriever, but the confidence score decreased to 0.49. After the second interaction, all parts were masked except for the face of the golden retriever. The classification result was changed to labrador retriever, with a confidence score of 0.83. These interactions demonstrate that the golden retriever's overall body and the surrounding background can influence the classification results.


For the soccer ball image, all parts except the ball were masked in the first interaction. The classification result was still a soccer ball, with a confidence score of 0.94. After the second interaction, only the soccer ball and the leg in contact with it were left unmasked. The classification result was a soccer ball, and the confidence score slightly decreased to 0.86. This shows that the object is the most critical feature for classifying the soccer ball image.


In the case of the coffee mug image, every part except for the mug itself was masked during the first interaction. The classification result was a coffee mug, with the confidence score increasing to 0.74. During the second interaction, the handle of the mug was also masked. Subsequently, the classification result changed to a pill bottle, with a confidence score of 0.44. These results indicate that the coffee mug, particularly its handle, significantly influences the model's output.


For the bakery image, everything except the shelves was masked in the first interaction. The classification result was changed to a tobacco shop, with a confidence score of 0.08. In the second interaction, the shelves, a person, and some interior elements remained unmasked. The masked image was classified as a bakery with a low confidence score of 0.46. This indicates that the classification model can be influenced not only by the primary object (e.g., shelves) but also by the surrounding interior or environment.


In the case of the cinema image, the audience seats were masked in the first interaction, leaving only the screen and stage. The classification result was a spotlight, with a confidence score of 0.14. In the second interaction, we unmasked some of the audience seats. The classification result then shifted to a cinema, with a lower confidence score of 0.32. This demonstrates that for cinema images, not only the screen but also the seats play crucial roles in image classification.


\section{Conclusions}

In this paper, we present an interaction-based xAI method tailored for image classification tasks. The method facilitates the generation of masked images through user interactions, allowing users to understand the model by examining the classification results of these masked images. We conducted experiments using five image samples to observe the changes in image classification results through the proposed method. The experiments demonstrated that classification results could vary according to the masked features and the importance of features. This shows that the interaction-based xAI method can help users comprehend which features are important for classifying images with computer vision models. 

However, our work has several limitations. First, the method was applied solely to a basic image classification model. Given that computer vision models encompass a wide range of tasks, it is crucial to design interactions tailored to each specific task. Furthermore, our work focused on basic interactions, but exploring other interaction types, such as pinch, zoom, and tap, could be beneficial. Moreover, future considerations might include not only the creation of masked images but also modifications to the current image. Lastly, our research did not involve a user study; we only observed the changes in classification results from sample images. To ascertain whether there is an improvement in users' understanding of computer vision models, conducting a user study is essential. Consequently, our future work aims to design a broader range of interactions, consider various computer vision tasks and models, and conduct a rigorous user study. Furthermore, we consider the approach that integrates our method with existing xAI techniques to provide a combination of automatic explanations and additional interactions, potentially reducing the challenges associated with iteration in the interaction-based method.  Despite the limitations, our method is significant for proposing an xAI method that leverages end-users' interactions.

\bibliographystyle{named}
\bibliography{ijcai24}

\begin{thebibliography}{}

\bibitem[\protect\citeauthoryear{Arkin \bgroup \em et al.\egroup }{2021}]{arkin2021survey}
Ershat Arkin, Nurbiya Yadikar, Yusnur Muhtar, and Kurban Ubul.
\newblock A survey of object detection based on cnn and transformer.
\newblock In {\em 2021 IEEE 2nd international conference on pattern recognition and machine learning (PRML)}, pages 99--108. IEEE, 2021.

\bibitem[\protect\citeauthoryear{Deng \bgroup \em et al.\egroup }{2009}]{deng2009imagenet}
Jia Deng, Wei Dong, Richard Socher, Li-Jia Li, Kai Li, and Li~Fei-Fei.
\newblock Imagenet: A large-scale hierarchical image database.
\newblock In {\em 2009 IEEE conference on computer vision and pattern recognition}, pages 248--255. Ieee, 2009.

\bibitem[\protect\citeauthoryear{Du \bgroup \em et al.\egroup }{2019}]{du2019eventaction}
Fan Du, Catherine Plaisant, Neil Spring, Kenyon Crowley, and Ben Shneiderman.
\newblock Eventaction: A visual analytics approach to explainable recommendation for event sequences.
\newblock {\em ACM Transactions on Interactive Intelligent Systems (TiiS)}, 9(4):1--31, 2019.

\bibitem[\protect\citeauthoryear{Dwivedi \bgroup \em et al.\egroup }{2023}]{dwivedi2023explainable}
Rudresh Dwivedi, Devam Dave, Het Naik, Smiti Singhal, Rana Omer, Pankesh Patel, Bin Qian, Zhenyu Wen, Tejal Shah, Graham Morgan, et~al.
\newblock Explainable ai (xai): Core ideas, techniques, and solutions.
\newblock {\em ACM Computing Surveys}, 55(9):1--33, 2023.

\bibitem[\protect\citeauthoryear{Elngar \bgroup \em et al.\egroup }{2021}]{elngar2021image}
Ahmed~A Elngar, Mohamed Arafa, Amar Fathy, Basma Moustafa, Omar Mahmoud, Mohamed Shaban, and Nehal Fawzy.
\newblock Image classification based on cnn: a survey.
\newblock {\em Journal of Cybersecurity and Information Management}, 6(1):18--50, 2021.

\bibitem[\protect\citeauthoryear{Goldstein \bgroup \em et al.\egroup }{2015}]{goldstein2015peeking}
Alex Goldstein, Adam Kapelner, Justin Bleich, and Emil Pitkin.
\newblock Peeking inside the black box: Visualizing statistical learning with plots of individual conditional expectation.
\newblock {\em journal of Computational and Graphical Statistics}, 24(1):44--65, 2015.

\bibitem[\protect\citeauthoryear{He \bgroup \em et al.\egroup }{2016}]{he2016deep}
Kaiming He, Xiangyu Zhang, Shaoqing Ren, and Jian Sun.
\newblock Deep residual learning for image recognition.
\newblock In {\em Proceedings of the IEEE conference on computer vision and pattern recognition}, pages 770--778, 2016.

\bibitem[\protect\citeauthoryear{Jiang \bgroup \em et al.\egroup }{2023}]{jiang2023situation}
Jinglu Jiang, Alexander~J Karran, Constantinos~K Coursaris, Pierre-Majorique L{\'e}ger, and Joerg Beringer.
\newblock A situation awareness perspective on human-ai interaction: Tensions and opportunities.
\newblock {\em International Journal of Human--Computer Interaction}, 39(9):1789--1806, 2023.

\bibitem[\protect\citeauthoryear{Khan \bgroup \em et al.\egroup }{2022}]{khan2022transformers}
Salman Khan, Muzammal Naseer, Munawar Hayat, Syed~Waqas Zamir, Fahad~Shahbaz Khan, and Mubarak Shah.
\newblock Transformers in vision: A survey.
\newblock {\em ACM computing surveys (CSUR)}, 54(10s):1--41, 2022.

\bibitem[\protect\citeauthoryear{Khater \bgroup \em et al.\egroup }{2023}]{khater2023skin}
Tarek Khater, Sam Ansari, Soliman Mahmoud, Abir Hussain, and Hissam Tawfik.
\newblock Skin cancer classification using explainable artificial intelligence on pre-extracted image features.
\newblock {\em Intelligent Systems with Applications}, 20:200275, 2023.

\bibitem[\protect\citeauthoryear{Kim \bgroup \em et al.\egroup }{2018}]{kim2018interpretability}
Been Kim, Martin Wattenberg, Justin Gilmer, Carrie Cai, James Wexler, Fernanda Viegas, et~al.
\newblock Interpretability beyond feature attribution: Quantitative testing with concept activation vectors (tcav).
\newblock In {\em International conference on machine learning}, pages 2668--2677. PMLR, 2018.

\bibitem[\protect\citeauthoryear{Kim \bgroup \em et al.\egroup }{2023}]{kim2023help}
Sunnie~SY Kim, Elizabeth~Anne Watkins, Olga Russakovsky, Ruth Fong, and Andr{\'e}s Monroy-Hern{\'a}ndez.
\newblock "help me help the ai": Understanding how explainability can support human-ai interaction.
\newblock In {\em Proceedings of the 2023 CHI Conference on Human Factors in Computing Systems}, pages 1--17, 2023.

\bibitem[\protect\citeauthoryear{Liao and Varshney}{2021}]{liao2021human}
Q~Vera Liao and Kush~R Varshney.
\newblock Human-centered explainable ai (xai): From algorithms to user experiences.
\newblock {\em arXiv preprint arXiv:2110.10790}, 2021.

\bibitem[\protect\citeauthoryear{Liao \bgroup \em et al.\egroup }{2020}]{liao2020questioning}
Q~Vera Liao, Daniel Gruen, and Sarah Miller.
\newblock Questioning the ai: informing design practices for explainable ai user experiences.
\newblock In {\em Proceedings of the 2020 CHI conference on human factors in computing systems}, pages 1--15, 2020.

\bibitem[\protect\citeauthoryear{Mothilal \bgroup \em et al.\egroup }{2020}]{mothilal2020explaining}
Ramaravind~K Mothilal, Amit Sharma, and Chenhao Tan.
\newblock Explaining machine learning classifiers through diverse counterfactual explanations.
\newblock In {\em Proceedings of the 2020 conference on fairness, accountability, and transparency}, pages 607--617, 2020.

\bibitem[\protect\citeauthoryear{Narwaria}{2022}]{narwaria2022does}
Manish Narwaria.
\newblock Does explainable machine learning uncover the black box in vision applications?
\newblock {\em Image and Vision Computing}, 118:104353, 2022.

\bibitem[\protect\citeauthoryear{Patr{\'\i}cio \bgroup \em et al.\egroup }{2023}]{patricio2023explainable}
Cristiano Patr{\'\i}cio, Jo{\~a}o~C Neves, and Lu{\'\i}s~F Teixeira.
\newblock Explainable deep learning methods in medical image classification: A survey.
\newblock {\em ACM Computing Surveys}, 56(4):1--41, 2023.

\bibitem[\protect\citeauthoryear{Recio-Garc{\'\i}a \bgroup \em et al.\egroup }{2021}]{recio2021case}
Juan~A Recio-Garc{\'\i}a, Humberto Parejas-Llanovarced, Mauricio~G Orozco-del Castillo, and Esteban~E Brito-Borges.
\newblock A case-based approach for the selection of explanation algorithms in image classification.
\newblock In {\em Case-Based Reasoning Research and Development: 29th International Conference, ICCBR 2021, Salamanca, Spain, September 13--16, 2021, Proceedings 29}, pages 186--200. Springer, 2021.

\bibitem[\protect\citeauthoryear{Ribeiro \bgroup \em et al.\egroup }{2016}]{ribeiro2016should}
Marco~Tulio Ribeiro, Sameer Singh, and Carlos Guestrin.
\newblock " why should i trust you?" explaining the predictions of any classifier.
\newblock In {\em Proceedings of the 22nd ACM SIGKDD international conference on knowledge discovery and data mining}, pages 1135--1144, 2016.

\bibitem[\protect\citeauthoryear{Selvaraju \bgroup \em et al.\egroup }{2017}]{selvaraju2017grad}
Ramprasaath~R Selvaraju, Michael Cogswell, Abhishek Das, Ramakrishna Vedantam, Devi Parikh, and Dhruv Batra.
\newblock Grad-cam: Visual explanations from deep networks via gradient-based localization.
\newblock In {\em Proceedings of the IEEE international conference on computer vision}, pages 618--626, 2017.

\bibitem[\protect\citeauthoryear{Shneiderman}{2020}]{shneiderman2020bridging}
Ben Shneiderman.
\newblock Bridging the gap between ethics and practice: guidelines for reliable, safe, and trustworthy human-centered ai systems.
\newblock {\em ACM Transactions on Interactive Intelligent Systems (TiiS)}, 10(4):1--31, 2020.

\bibitem[\protect\citeauthoryear{Steiner \bgroup \em et al.\egroup }{2019}]{steiner2019pytorch}
Benoit Steiner, Zachary DeVito, Soumith Chintala, Sam Gross, Adam Paske, Francisco Massa, Adam Lerer, Greg Chanan, Zeming Lin, Edward Yang, et~al.
\newblock Pytorch: An imperative style, high-performance deep learning library.
\newblock 2019.

\bibitem[\protect\citeauthoryear{Von~Eschenbach}{2021}]{von2021transparency}
Warren~J Von~Eschenbach.
\newblock Transparency and the black box problem: Why we do not trust ai.
\newblock {\em Philosophy \& Technology}, 34(4):1607--1622, 2021.

\bibitem[\protect\citeauthoryear{Weld and Bansal}{2019}]{weld2019challenge}
Daniel~S Weld and Gagan Bansal.
\newblock The challenge of crafting intelligible intelligence.
\newblock {\em Communications of the ACM}, 62(6):70--79, 2019.

\bibitem[\protect\citeauthoryear{Yeh \bgroup \em et al.\egroup }{2018}]{yeh2018representer}
Chih-Kuan Yeh, Joon Kim, Ian En-Hsu Yen, and Pradeep~K Ravikumar.
\newblock Representer point selection for explaining deep neural networks.
\newblock {\em Advances in neural information processing systems}, 31, 2018.

\bibitem[\protect\citeauthoryear{Yun \bgroup \em et al.\egroup }{2023}]{yun2023driving}
Hyeonggeun Yun, Younggeol Cho, Arim Ha, and Jihyeok Yun.
\newblock Driving with black box assistance: Teleoperated driving interface design guidelines for computational driver assistance systems in unstructured environments.
\newblock In {\em Proceedings of the 15th International Conference on Automotive User Interfaces and Interactive Vehicular Applications}, pages 156--166, 2023.

\end{thebibliography}

\end{document}